\documentclass[12pt]{article}

\begin{document}

\begin{center}
{\bf On First-Order Generalized Maxwell Equations } \\
\vspace{5mm} S. I. Kruglov \footnote{E-mail:
krouglov@utsc.utoronto.ca}\\
 \vspace{3mm} \textit{University of Toronto at Scarborough,\\ Physical
and Environmental Sciences Department, \\
1265 Military Trail, Toronto, Ontario, Canada M1C 1A4} \\

\vspace{5mm}
\end{center}

\begin{abstract}
The generalized Maxwell equations including an additional scalar
field are considered in the first-order formalism. The gauge
invariance of the Lagrangian and equations is broken resulting in
the appearance of a scalar field. We find the canonical and
symmetrical Belinfante energy-momentum tensors. It is shown that
the traces of the energy-momentum tensors are not equal to zero
and the dilatation symmetry is broken in the theory considered.
The matrix Hamiltonian form of equations is obtained after the
exclusion of the nondynamical components. The canonical
quantization is performed and the propagator of the fields is
found in the first-order formalism.
\end{abstract}

\section{Introduction}

In \cite{Kruglov} we found and investigated the matrix
(first-order) formulation of the generalized Maxwell equations,
and solutions for a free particle in the form of the projection
matrix-dyads. The matrices of an equation obey the generalized
Duffin-Kemmer algebra. The generalized Maxwell equations can be
obtained by breaking the gauge invariance and refusing the Lorentz
condition, $\partial_\mu A_\mu\neq 0$. As a result, a scalar field
$\varphi=\partial_\mu A_\mu$ appears because the vector field
realizes the $(1/2,1/2)$ representation of the Lorenz group and
possesses four degrees of freedom (three of them describe spin one
and another $-$ spin zero). The scalar field here is a ghost field
which can be removed by the path integration formulation of
electrodynamics \cite{Faddeev}. The reason why we leave this ghost
field is scalar fields play an important role in the Standard
Model (the Higgs field) and in astrophysics (the inflation field).
It should be noted that in some scenarios of Universe inflation
ghost scalar fields (k-essence, Phantom) are insensitively
explored \cite{Mukhanov}. An interesting scenario of inflation
deals with phenomenon of the ghost condensation
\cite{Arkani-Hamed}.

In this work we find the canonical and symmetrical Belinfante
energy-momentum tensors, the dilatation current, and study the
matrix Hamilton form of equations, and the canonical quantization.

The paper is organized as follows. In Sec.2, we formulate the
generalized Maxwell equations in the matrix form. The canonical
and the symmetrical Belinfante energy momentum tensors are
obtained in Sec.3. We find the dilatation current and demonstrate
that the dilatation symmetry is broken. The quantum-mechanical
Hamiltonian is obtained in Sec.4. In Sec.5, the canonical
quantization of fields is performed and the matrix propagator of
fields is obtained. We draw a conclusion in Sec.6. The Euclidean
metric is used and the system of units $\hbar =c=1$ is chosen.

\section{Matrix form of equations}

The generalized Maxwell equations may be cast as follows
\cite{Kruglov}:
\[
\partial _\nu \psi _{[\mu \nu ]}+\partial _\mu \psi _0=0,
\]
\begin{equation}
\partial _\nu \psi _\mu -\partial _\mu \psi _\nu +\kappa\psi _{[\mu \nu ]}=0,
\end{equation}
\[
\partial _\mu \psi_\mu +\kappa\psi _0=0.  \label{1}
\]
The mass parameter $\kappa$ is introduced to have fields $\psi_A$
($A=0$, $\mu$, $[\mu\nu]$) with the same dimension. Physical
values do not depend on $\kappa$. It follows from Eq.(1) that
fields $\psi_A$ satisfy the wave equation:
$\partial_\mu^2\psi_A(x)=0$ that does not depend on the
dimensional parameter $\kappa$. In the second-order formulation of
the theory, the fields $\psi_0$ and $\psi_{[\mu\nu]}$ can be
eliminated and all that is left is the field $\psi_\mu$ obeying
the wave equation. The scope of our consideration is the
first-order formulation of the theory and, therefore, we introduce
the auxiliary fields $\psi_0$ and $\psi_{[\mu\nu]}$. One can treat
Eq.(1) as the Maxwell equations in the general gauge
\cite{Faddeev}. The gradient term $j_\mu=-\partial _\mu \psi _0$
plays the role of the four-current and depends on the
vector-potential $\psi _\mu$. In such electrodynamics the
vector-potential $\psi _\mu$ is a ``physical" variable as well as
the strengths of the ``electric", $E_m=i\psi _{[m 4]}$, and
``magnetic", $H_k=(1/2)\varepsilon_{kmn}\psi _{[m n]}$, fields.

Eq.(1) can be written in the $11\times11$-matrix form
\cite{Kruglov}:
\begin{equation}
\left( \alpha _\nu \partial _\nu +\kappa P\right) \Psi (x)=0,
\label{2}
\end{equation}
where
\[
\alpha _\mu =\beta _\mu ^{(1)}+\beta _\mu ^{(0)},~~\beta _\mu
^{(1)}=\varepsilon ^{\nu ,[\nu \mu ]}+\varepsilon ^{[\nu \mu ],\nu
},~~\beta _\mu ^{(0)}=\varepsilon ^{\mu ,0}+\varepsilon ^{0,\mu },
\]
\vspace{-8mm}
\begin{equation}
\label{3}
\end{equation}
\vspace{-8mm}
\[
P=\varepsilon ^{0,0}+\frac 12\varepsilon ^{[\mu \nu ],[\mu \nu
]},~~~\Psi (x)=\left\{ \psi _A(x)\right\},
\]
and elements of the entire matrix algebra $\varepsilon ^{A,B}$
obey equations: $\left( \varepsilon
^{A,B}\right)_{CD}=\delta_{AC}\delta_{BD}$, $\varepsilon
^{A,B}\varepsilon ^{C,D}=\delta _{BC}\varepsilon ^{A,D}$. Matrices
$\alpha _\nu $, $P$ are Hermitian matrices, and the $P$ is the
projection operator, $P^2=P$.

The $11\times11$-matrices $\alpha_\mu$ satisfy the algebra as
follows \cite{Kruglov}:
\[
\alpha _\mu \alpha _\nu \alpha _\alpha +\alpha _\alpha \alpha _\nu
\alpha _\mu +\alpha _\mu \alpha _\alpha \alpha _\nu +\alpha _\nu
\alpha _\alpha \alpha _\mu +\alpha _\nu \alpha _\mu \alpha _\alpha
+\alpha _\alpha \alpha _\mu \alpha _\nu =
\]
\vspace{-8mm}
\begin{equation}
\label{4}
\end{equation}
\vspace{-8mm}
\[
 =2\left( \delta _{\mu \nu }\alpha _\alpha +\delta
_{\alpha \nu }\alpha _\mu +\delta _{\mu \alpha }\alpha _\nu
\right).
\]

The Lagrangian of the theory suggested is given by
\begin{equation}
{\cal L}=-\overline{\Psi }(x)\left(\alpha _\mu
\partial _\mu +\kappa P\right) \Psi (x),
\label{5}
\end{equation}
where $ \overline{\Psi }(x)\Psi (x)=\Psi ^{+}(x)\eta \Psi (x)$
is the Lorentz invariant, and the Hermitianizing matrix $\eta$ is
\begin{equation}
\eta =-\varepsilon ^{0,0}+\varepsilon ^{m,m}-\varepsilon ^{4,4}
+\varepsilon ^{[m4],[m4]}-\frac{1}{2}\varepsilon ^{[mn],[mn]}.
\label{6}
\end{equation}
The $\eta $ is the Hermitian matrix, $\eta^+ =\eta $, and the
Lagrangian (5) is the real function that can be verified with the
help of relations: $\eta \alpha _m=-\alpha _m^{+}\eta^+$
(m=1,2,3), $\eta \alpha _4=\alpha _4^{+}\eta^+ $. In terms of
fields $\psi_A$, the Lagrangian (5) reads
\begin{equation}
{\cal L}=\psi_0 \partial _\mu\psi_\mu -\psi_\mu\partial _\mu
\psi_0 -\psi_\rho\partial _\mu\psi_{[\rho\mu]}+
\psi_{[\rho\mu]}\partial _\mu\psi_\rho +\kappa\left(\psi_0^2
+\frac{1}{2}\psi_{[\rho\mu]}^2\right). \label{7}
\end{equation}
It is easy to verify that Euler-Lagrange equations $\partial{\cal
L}/\partial\psi_A-\partial_\mu\left(\partial{\cal L}/
\partial\partial_\mu\psi_A\right)=0$ lead to Eq.(1). For fields
$\psi_A$ obeying Eq.(1), the Lagrangian (7) vanishes.
We can exclude the auxiliary fields $\psi_0$ and $\psi_{[\mu\nu]}$
from Eq.(7), and as a result, the Lagrangian reads
\begin{equation}
{\cal L}=\frac{1}{\kappa}\psi_\mu\partial_\alpha^2 \psi_\mu. \label{8}
\end{equation}
To have the standard kinetic term, one may renormalize the fields, and within
the 4-divergence, the Lagrangian becomes
\begin{equation}
{\cal L'}=-\frac{1}{2}\left(\partial_\nu A_\mu\right)^2. \label{9}
\end{equation}
The second-order formulation based on the simple Lagrangian (9)
leads to the negative metric state \cite{Kruglov2},
\cite{Kruglov1}. This occurs because of the presence of the ghost.
In the theory with the gage invariance, the negative metric state
can be decoupled. One can exclude the negative metric state by
putting $\psi_0=0$ in the model considered.

\section{The energy-momentum tensor}

The canonical energy-momentum tensor can be found with the help of
the standard procedure, and is given by
\begin{equation}
T^c_{\mu\nu}=\left(\partial_\nu \overline{\Psi} (x)\right)\alpha_\mu
\Psi (x). \label{10}
\end{equation}
With the help of Eq.(3) it becomes
\begin{equation}
T^c_{\mu\nu}=\psi_0\partial_\nu \psi_\mu-\psi_\mu\partial_\nu
\psi_0-\psi_\rho\partial_\nu \psi_{[\rho\mu]} +\psi_{[\rho\mu]}
\partial_\nu \psi_\rho. \label{11}
\end{equation}
One can verify, using Eq.(1), that the energy-momentum tensor (11)
(or (10)) is conserved tensor, $\partial_\mu T^c_{\mu\nu}=0$, but it
is not the symmetric tensor, $T^c_{\mu\nu}\neq T^c_{\nu\mu}$. The
symmetric energy-momentum tensor can be found from the relation
\cite{Landau}
\begin{equation}
T_{\mu\nu}^{sym}=T_{\mu\nu}+\Lambda_{\mu\nu}, \label{12}
\end{equation}
where the function $\Lambda_{\mu\nu}$ has to satisfy the equation
$\partial_\mu \Lambda_{\mu\nu}=0$ for the fields $\psi_A$ obeying
Eq.(1) in such a way that $T_{\mu\nu}^{sym}=T_{\nu\mu}^{sym}$. We
obtain
\begin{equation}
\Lambda_{\mu\nu}=\psi_0\partial_\mu \psi_\nu-\psi_\nu\partial_\mu
\psi_0-\psi_\rho\partial_\mu \psi_{[\rho\nu]} +\psi_{[\rho\nu]}
\partial_\mu \psi_\rho. \label{13}
\end{equation}
It is interesting that contrary to the classical electrodynamics,
the trace of the symmetric energy-momentum tensor (12) is not
equal to zero
\begin{equation}
T_{\mu\mu}^{sym}=-\kappa\left(2\psi_0^2 +\psi_{[\mu\nu]}^2\right).
\label{14}
\end{equation}
Usually, the nonzero trace of the energy-momentum tensor appears in the massive
theories with broken dilatation and conformal symmetries.

To investigate the dilatation symmetry, we consider the canonical
dilatation current \cite{Coleman}
\begin{equation}
D_\mu^c=x_\alpha T_{\mu\alpha}^{c}+\Pi_\mu \Psi,
\label{15}
\end{equation}
where
\begin{equation}
\Pi_\mu=\frac{{\partial\cal L}}{\partial\left(\partial_\mu\Psi\right)}
=-\overline{\Psi }\alpha_\mu,
\label{16}
\end{equation}
and we took into account that for the Bose fields, the matrix $\textbf{d}$ defining
the field dimension, is the unit matrix. It is easy to verify, with the help of the
expression $\overline{\Psi }=\left(-\psi_0,\psi_\mu,-\psi_{[\mu\nu]}\right)$ and
Eq.(3), that the second term in
Eq.(15) is the electric current $J_\mu$, and it vanishes:
\begin{equation}
J_\mu=\Pi_\mu \Psi=-\overline{\Psi }\alpha_\mu\Psi=0.
\label{17}
\end{equation}
It is obvious that for neutral fields the electric current should be zero.
From Eq.(15), one finds
\begin{equation}
\partial_\mu D_\mu^c=T_{\mu\mu}^{c}=-\kappa\left(\psi_0^2 +
\frac{1}{2}\psi_{[\mu\nu]}^2\right).
\label{18}
\end{equation}
The same expression may be obtained from the relation \cite{Coleman}
\begin{equation}
\partial_\mu D_\mu^c=2\Pi_\mu\partial_\mu\Psi +\frac{\partial{\cal L}}{\partial\Psi}\Psi -4{\cal L}
=\kappa\overline{\Psi }P\Psi=-\kappa\left(\psi_0^2 +
\frac{1}{2}\psi_{[\mu\nu]}^2\right).
\label{19}
\end{equation}
Thus, the dilatation symmetry is broken in spite of the fields
being massless. This occurs because of the presence of the
parameter $\kappa$ with the dimension of the mass. It should be
noted, that the necessity of the dimensional parameter is because
that components of the field $\Psi$ ($\psi_0$, $\psi_\mu$,
$\psi_{[\mu\nu]}$) should have the same dimension. As a result,
the first-order formulation of massless fields breaks the
dilatation symmetry. The appearance of mass parameters in the full
theory usually takes place due to condensations of fields, and as
a result, the vacuum expectations of fields are nonzero. In this
toy model of free fields, we do not specify the nature of the
parameter $\kappa$.

We notice that the procedure of the energy-momentum symmetrization
can be done in different ways \cite{Landau}. It is fruitful to
obtain the Belinfante tensor \cite{Coleman}
 \begin{equation}
T_{\mu\alpha}^{B}=T_{\mu\alpha}^{c}+ \partial_\beta X_{\beta\mu\alpha},
\label{20}
\end{equation}
where
\begin{equation}
X_{\beta\mu\alpha}=\frac{1}{2}\left[\Pi_\beta J_{\mu\alpha}\Psi-
\Pi_\mu J_{\beta\alpha}\Psi-\Pi_\alpha J_{\beta\mu}\Psi\right].
\label{21}
\end{equation}
The generators of the Lorentz group in the 11-dimensional representation
space are given by \cite{Kruglov}
\begin{equation}
J_{\mu\alpha}=\beta^{(1)}_\mu\beta^{(1)}_\alpha-\beta^{(1)}_\alpha\beta^{(1)}_\mu
=\varepsilon^{\mu,\alpha}-\varepsilon^{\alpha,\mu}+\varepsilon^{[\lambda\mu],[\lambda\alpha]}
-\varepsilon^{[\lambda\alpha],[\lambda\mu]}.
\label{22}
\end{equation}
It is easy to obtain the tensor $X_{\beta\mu\alpha}$ from Eq.(21)
in terms of components $\psi_A$:
\begin{equation}
X_{\beta\mu\alpha}=\delta_{\alpha\mu}\psi_0\psi_\beta-\delta_{\alpha\beta}\psi_0\psi_\mu+
\delta_{\alpha\beta}\psi_\lambda\psi_{[\lambda\mu]}-\delta_{\alpha\mu}\psi_\lambda\psi_{[\lambda\beta]}
-2\psi_\alpha\psi_{[\beta\mu]}.
\label{23}
\end{equation}
Replacing the expression (23) into Eq.(20), one finds the Belinfante
symmetric energy-momentum tensor
\[
T_{\mu\alpha}^{B}=2\kappa\psi_{[\lambda\mu]}\psi_{[\alpha\lambda]}-2\psi_\mu\partial_\alpha\psi_0
-2\psi_\alpha\partial_\mu\psi_0
\]
\vspace{-8mm}
\begin{equation}
\label{24}
\end{equation}
\vspace{-8mm}
\[
+\delta_{\alpha\mu}\partial_\beta\left(\psi_0\psi_\beta\right)
-\delta_{\alpha\mu}\partial_\beta\left(\psi_\lambda\psi_{[\lambda\beta]}\right).
\]
 Now we can evaluate a modified dilatation current \cite{Coleman}
 \begin{equation}
D_\mu^B=x_\alpha T_{\mu\alpha}^{B}+V_\mu,
\label{25}
\end{equation}
where the field-virial $V_\mu$ in our case becomes
 \begin{equation}
V_\mu=\Pi_\mu \Psi-\Pi_\alpha J_{\alpha\mu}\Psi=\overline{\Psi }\alpha_\alpha
J_{\alpha\mu}\Psi=\psi_\lambda\psi_{[\lambda\mu]}-3\psi_0\psi_\mu.
\label{26}
\end{equation}
Calculating the divergence of the Belinfante dilatation current (25), one obtains
\begin{equation}
\partial_\mu D_\mu^B=T_{\mu\mu}^{B}+\partial_\mu V_\mu=-\kappa\left(\psi_0^2 +
\frac{1}{2}\psi_{[\mu\nu]}^2\right).
\label{27}
\end{equation}
Thus, we came to the same result (see Eq.(19)) that the dilatation current is not
conserved, i.e. the dilatation symmetry is broken. The conformal invariance is
also broken because the field-virial $V_\mu$ (see Eq.26) is not a total derivative
of some local quantity $\sigma_{\alpha\beta}$ \cite{Coleman}.

Now we turn to the second order formulation based on the simple Lagrangian (9). This
Lagrangian describes massless fields with spin-1 and spin-0 states obeying the
equation of motion: $\partial^2_\alpha A_\mu=0$. The canonical energy-momentum tensor
becomes
\begin{equation}
\Theta^c_{\mu\alpha}=\frac{\partial {\cal L'}}{\partial\left(\partial_\mu A_\beta\right)}
\partial_\alpha A_\beta-\delta_{\mu\alpha}{\cal L'}=
-\left(\partial_\mu A_\beta\right)\left(\partial_\alpha A_\beta\right)
+\frac{1}{2}\delta_{\mu\alpha}\left(\partial_\nu A_\beta\right)^2,
\label{28}
\end{equation}
and is also the symmetrical tensor. It easy to check, using the wave equation,
that the canonical energy-momentum tensor is conserved: $\partial_\mu \Theta^c_{\mu\alpha}=0$.
From Eq.(28) one finds the non-zero energy-momentum trace
\begin{equation}
\Theta^c_{\mu\mu}=\left(\partial_\mu A_\beta\right)^2.
\label{29}
\end{equation}
The dilatation current is given by
\begin{equation}
D^c_{\mu}=x_\alpha \Theta^c_{\mu\alpha}
+\frac{\partial {\cal L'}}{\partial\left(\partial_\mu A_\beta\right)}A_\beta=
x_\alpha \Theta^c_{\mu\alpha}-A_\beta\partial_\mu A_\beta.
\label{30}
\end{equation}
One can verify with the help of Eq.(29) that the dilatation
current (30) is conserved: $\partial_\mu D^c_{\mu}=0$, i.e. the
model based on the Lagrangian (9) possesses the scale invariance.
Thus, in spite of non-zero trace (29), the dilatation symmetry
takes place because there is no dimensional parameter in the
theory. It should be noted that the Belinfante energy-momentum
tensor is different from the symmetrical energy-momentum tensor
(28). Simple calculations show that the modified Belinfante
dilatation current is conserved. This confirms the dilatation
invariance of the model based on the Lagrangian (9).

\section{Quantum-mechanical Hamiltonian}

The quantum-mechanical Hamiltonian may be obtained from Eq.(2).
From Eq.(2), we arrive at
\begin{equation}
i\alpha _4\partial _t\Psi (x)=\left(\alpha _a\partial_a+\kappa
P\right)\Psi (x).
 \label{31}
\end{equation}
One obtains from algebra (4) the relation as follows:
\begin{equation}
\alpha _4\left(\alpha _4^2-1 \right)=0.
 \label{32}
\end{equation}
Eq.(32) indicates that eigenvalues of the matrix $\alpha _4$ are
one and zero. As a result, the 11-dimensional function $\Psi (x)$
possesses as dynamical as well as non-dynamical components.

To separate the dynamical and non-dynamical components of the wave
function, we introduce projection operators:
\begin{equation}
\Lambda\equiv\alpha_4^2=\varepsilon^{0,0}+\varepsilon^{\mu,\mu}+
\varepsilon^{[m4],[m4]},~~~~\Pi \equiv 1-\Lambda
=\frac{1}{2}\varepsilon^{[mn],[mn]}, \label{33}
\end{equation}
so that $\Lambda=\Lambda^2$, $\Pi^2=\Pi$, $\Lambda\Pi=0$. These
operators extract dynamical, $\phi(x)$, and non-dynamical
$\chi(x)$ components:
\begin{equation}
\phi(x)=\Lambda\Psi(x),~~~~\chi(x)=\Pi \Psi(x). \label{34}
\end{equation}
Multiplying Eq.(31) by the matrix $\alpha_4$ and $\Pi$, one
obtains the system of equations
\begin{equation}
i\partial _t \phi(x)=\alpha_4\left(\alpha_a \partial_a+\kappa P
\right)\left(\phi(x)+\chi(x)\right), \label{35}
\end{equation}
\begin{equation}
0=\left(\alpha _4^2-1 \right)\left(\alpha_a \partial_a+\kappa P
\right)\left(\phi(x)+\chi(x)\right) \label{36}
\end{equation}
With the help of the equation $P \Pi =\Pi P=\Pi$, one obtains from
Eq.(36), the expression
\begin{equation}
\chi(x)=\frac{1}{\kappa}\left(\alpha _4^2-1 \right)\alpha_a
\partial_a\phi(x) . \label{37}
\end{equation}
Replacing the $\chi(x)$ from Eq.(37) into Eq.(35), with the aid of
the equality $\alpha_4\Pi=0$, we find the Hamiltonian form:
\[
i\partial _t\phi (x)=\mathcal{H}\phi (x) ,
\]
\vspace{-7mm}
\begin{equation} \label{38}
\end{equation}
\vspace{-7mm}
\[
\mathcal{H}=\alpha_4\left(\alpha_a \partial_a+\kappa P \right)
-\frac{1}{\kappa}\alpha_4\alpha_a\Pi\alpha_b\partial_a\partial_b.
\]
The $8$-component Eq.(38) describes four spin states (spin one and
zero) with positive and negative energies. In the component form
Eq.(38) leads to equations as follows:
\[
i\partial_t\psi_0=\partial_m\psi_{[4m]},~~~~i\partial_t\psi_m=
-\partial_m\psi_4+\kappa\psi_{[m4]},
\]
\vspace{-7mm}
\begin{equation} \label{39}
\end{equation}
\vspace{-7mm}
\[
i\partial_t\psi_4=
\partial_m\psi_m+\kappa\psi_0,~~~~i\partial_t\psi_{[m4]}=
\partial_m\psi_0+\frac{1}{\kappa}\left(\partial_m\partial_n\psi_n
-\partial_n^2\psi_m\right).
\]
Eq.(39) can also be obtained from Eq.(1) retaining only
components with time derivatives. The Hamiltonian $\mathcal{H}$ describes
the evolution of the wave function $\phi(x)$ in time. Eq.(37) is
equivalent to the equation
$\kappa\psi_{[mn]}=\partial_m\psi_n-\partial_n\psi_m$ which does
not contain the time derivative.

\section{ Field Quantization}

We obtain from Eq.(5) the momenta
\begin{equation}
\pi (x)=\frac{\partial\mathcal{L}}{\partial(\partial_0\Psi
(x))}=i\overline{\Psi}\alpha_4.
 \label{40}
\end{equation}
With the help of the relationship $[\Psi_M(\textbf{x},t),\pi_N
(\textbf{y},t)]= i \delta_{MN} \delta(\textbf{x}-\textbf{y})$, one
arrives from Eq.(40) at simultaneous quantum commutators
\begin{equation}
\left[\Psi_M(\textbf{x},t),
\left(\overline{\Psi}(\textbf{y},t)\alpha_4 \right)_N\right ]=
\delta_{MN}\delta(\textbf{x}-\textbf{y}) .\label{41}
\end{equation}
From Eq.(41), we obtain non-zero commutators of fields $\psi_A
(x)$:
\begin{equation}
\left[\psi_0(\textbf{x},t),\psi_4(\textbf{y},t)\right ]=
\delta(\textbf{x}-\textbf{y}),~~\left[\psi_{[m4]}(\textbf{x},t),\psi_n(\textbf{y},t)
\right]= \delta_{mn}\delta(\textbf{x}-\textbf{y}).\label{42}
\end{equation}

Let us consider solutions of Eq.(2) with definite energy and
momentum in the form of plane waves:
\begin{equation}
\Psi_\lambda^{(\pm)}(x)=\sqrt{\frac{1}{2k_0 V}}v_\lambda(\pm
k)\exp(\pm ikx) , \label{43}
\end{equation}
where $V$ is the normalization volume, $k^2=
\textbf{k}^2-k_0^2=0$, and $\lambda$ is the spin index
($\lambda=1,2,3,4$). The 11-dimensional function $v_\lambda(\pm
k)$ obeys the equation as follows:
\begin{equation}
\left(i\hat{k}\pm \kappa P \right)v_\lambda(\pm k)=0 , \label{44}
\end{equation}
where $\hat{k}=\alpha_\mu k_\mu$. We use the normalization
conditions
\begin{equation}
\int_V \overline{\Psi}^{(\pm)}_{\lambda}(x)\alpha_4
\Psi^{(\pm)}_{\lambda' }(x)d^3 x=\pm\delta_{\lambda\lambda'}
,~~~~\int_V \overline{\Psi}^{(\pm)}_{\lambda}(x)\alpha_4
\Psi^{(\mp)}_{\lambda' }(x)d^3 x=0 , \label{45}
\end{equation}
where
$\overline{\Psi}^{(\pm)}_{\lambda}(x)=\left(\Psi^{(\pm)}_{\lambda
}(x)\right)^+ \eta$. From normalization conditions (45), one finds
relations for functions $v_\lambda(\pm k)$:
\begin{equation}
\overline{v}_\lambda(\pm k)\alpha_\mu v_{\lambda'}(\pm k)=\mp
2ik_\mu\delta_{\lambda\lambda'} ,~~~~\overline{v}_\lambda(\pm
k)\alpha_4 v_{\lambda'}(\mp k)=0 . \label{46}
\end{equation}
Multiplying the first equation in (46) by $k_\mu$, one obtains
\begin{equation}
\overline{v}_\lambda(\pm k)\hat{k} v_{\lambda'}(\pm
k)=0,~~~~\overline{v}_\lambda(\pm k)P v_{\lambda'}(\pm k)=0 .
\label{47}
\end{equation}

The Hamiltonian density (energy density) is given by the equation
\begin{equation}
 {\cal E}=\pi (x)\partial_0\Psi (x) -{\cal L}
=i\overline{\Psi}(x)\alpha_4\partial_0\Psi(x).
 \label{48}
\end{equation}
In the second quantized theory, the field operators may be
represented as
\[
\Psi(x)=\sum_{k\lambda}\left[a_{k,\lambda}\Psi^{(+)}_{\lambda}(x)
+ a^+_{k,\lambda}\Psi^{(-)}_{\lambda}(x)\right] ,
\]
\vspace{-7mm}
\begin{equation} \label{49}
\end{equation}
\vspace{-7mm}
\[
\overline{\Psi}(x)=\sum_{k\lambda}\left[a^+_{k,\lambda}
\overline{\Psi}^{(+)}_{\lambda}(x)+ a_{k,\lambda}
\overline{\Psi}^{(-)}_{\lambda}(x)\right] ,
\]
where $a_{k,4}=ia_{k,0}$, $a^+_{k,4}=ia^+_{k,0}$, and the positive and negative
parts of the wave function are given
by Eq.(43). The creation and annihilation operators of particles,
$a^+_{k,\lambda}$, $a_{k,\lambda}$ obey the commutation relations
as follows:
\begin{equation}
[a_{k,\lambda},a^+_{k',\lambda'}]=\delta_{\lambda\lambda'}
\delta_{kk'} ,~~~
[a_{k,\lambda},a_{k',\lambda'}]=[a^+_{k,\lambda},a^+_{k',\lambda'}]=0.
\label{50}
\end{equation}
It should be noted that the operators $a_{k,0}$, $a^+_{k,0}$
satisfy the the commutation relation
$[a_{k,0},a^+_{k',0}]=-\delta_{kk'}$ with the ``wrong" sign ($-$).
With the aid of Eq.(48)-(50), and the normalization condition
(46), one obtains the Hamiltonian
\begin{equation}
H=\int {\cal E}d^3 x=\sum_{k,\lambda}k_0\left(a^+_{k,\lambda}
a_{k,\lambda}+a_{k,\lambda} a^+_{k,\lambda}\right) . \label{51}
\end{equation}
The classical Hamiltonian obtained from Eq.(5) or Eq.(9) is not bounded from below,
i.e. it is not positive-definite.
In the second quantized theory, the eigenvalues of the Hamiltonian (51) are positive if
we introduce the indefinite metric \cite{Kruglov2}, \cite{Kruglov1}. As a result, the square
norm of the ghost state is negative. This ghost state belongs to the ``nonphysical" Hilbert
subspace. But transitions between states from the physical subspace (with positive metric)
and the ghost violate the unitarity of the S-matrix and the probability interpretation. If we
impose the Lorentz condition $\partial_\mu\psi_\mu=0$ ($\psi_0=0$), only transverse components
will remain in the Hamiltonian (51) corresponding to two polarizations.

From Eq.(49)-(50), it is not difficult to find commutation
relations for different times
\begin{equation}
[\Psi_M(x),\Psi_N(x')]=[\overline{\Psi}_{ M}(x), \overline{\Psi}_{
N}(x')] =0, \label{52}
\end{equation}
\begin{equation}
[\Psi_{ M}(x),\overline{\Psi}_{N}(x')]=N_{ MN}(x,x'), \label{53}
\end{equation}
\[
N_{MN}(x,x')=N^+_{MN}(x,x')-N^-_{ MN}(x,x') ,
\]
\begin{equation}
N^+_{MN}(x,x')=\sum_{k,\lambda}\left(\Psi^{(+)}_{\lambda}(x)\right)_M
\left(\overline{\Psi}^{(+)}_{\lambda}(x')\right)_N  ,\label{54}
\end{equation}
\[
N^-_{MN}(x,x')=\sum_{k,\lambda}\left(\Psi^{(-)}_{\lambda}(x)\right)_M
\left(\overline{\Psi}^{(-)}_{\lambda}(x')\right)_N .
\]
We obtain from Eq.(54)
\begin{equation}
N^\pm_{MN}(x,x')=\sum_{k,\lambda}\frac{1}{2k_0
V}\left(v_{\lambda}(\pm
k)\right)_M\left(\overline{v}_{\lambda}(\pm k)\right)_N\exp [\pm
ik(x-x')] .
 \label{55}
\end{equation}
In \cite{Kruglov}, we found matrices-dyad for the spin-0 state
\begin{equation}
v^{(0)}_{0}(k)\cdot \overline{v}^{(0)}_{0}(k)=\left(
1-\frac{\sigma ^2}2\right) \left( \frac{i\hat{k}- \kappa
\overline{P} }{\kappa}\right) ^2,
 \label{56}
\end{equation}
and for spin-1 states\footnote{In \cite{Kruglov} we wrote out only
two solutions for spin-1 states (for helicity $\pm1$)}
\[
v_{\pm 1}(k)\cdot \overline{v}_{\pm 1}(k)=\frac 12\sigma _k\left(
\sigma _k\pm 1\right)\frac{\sigma ^2}2 \left( \frac{i\hat{k}-
\kappa \overline{P} }{\kappa}\right) ^2,
\]
\vspace{-6mm}
\begin{equation} \label{57}
\end{equation}
\vspace{-6mm}
\[
v_{0}(k)\cdot \overline{v}_{0}(k)= \left(1-\sigma _k^2\right)
\frac{\sigma ^2}2\left( \frac{i\hat{k}- \kappa \overline{P}
}{\kappa}\right) ^2 ,
\]
where $\sigma _k$ is the operator of the spin projection on the
direction of the momentum ${\bf k}$, $\sigma ^2$ is the square
spin operator, $\overline{P}=1-P$ is the projection operator, and
$(v\cdot \overline{v})_{MN}=v_{M}\overline{v}_N$. From Eq.(56),
(57), one obtains
\begin{equation}
\sum_\lambda\left(v_{\lambda}(\pm
k)\right)_M\left(\overline{v}_{\lambda}(\pm k)\right)_N =\left(
\frac{\pm i\hat{k}- \kappa \overline{P} }{\kappa}\right) ^2_{MN}.
\label{58}
\end{equation}
Taking into account Eq.(58), we find from Eq.(55):
\[
N^\pm_{MN}(x,x')=\sum_{k} \frac{1}{2k_0 V} \left( \frac{\pm
i\hat{k}- \kappa \overline{P} }{\kappa}\right) ^2_{MN} \exp [\pm
ik(x-x')]
\]
\vspace{-6mm}
\begin{equation} \label{59}
\end{equation}
\vspace{-6mm}
\[
= \left( \frac{ \alpha_\mu\partial_\mu- \kappa \overline{P}
}{\kappa}\right) ^2_{MN} D_{\pm}(x-x'),
\]
where the singular functions are given by \cite{Ahieser}
\begin{equation}
D_+(x)=\sum_{k}\frac{1}{2k_0V}\exp
(ikx),~~~~D_-(x)=\sum_{k}\frac{1}{2k_0V}\exp (-ikx). \label{60}
\end{equation}
With the help of the function \cite{Ahieser}
\begin{equation}
D_0 (x)=i\left(D_+(x)-D_-(x)\right), \label{61}
\end{equation}
from Eq.(54), (59), we obtain
\begin{equation}
N_{ MN}(x,x')=-i\left( \frac{\alpha_\mu\partial_\mu- \kappa
\overline{P} }{\kappa}\right) ^2_{MN} D_{0}(x-x').
 \label{62}
\end{equation}

It easy to verify, using Eq.(3), that the equation
\begin{equation}
 \left(\alpha_\mu\partial_\mu+ \kappa
P \right)\left( \frac{\alpha_\nu\partial_\nu- \kappa \overline{P}
}{\kappa}\right) ^2=\left( \frac{\alpha_\mu\partial_\mu- \kappa
\overline{P} }{\kappa^2}\right)\partial_\alpha^2
 \label{63}
\end{equation}
is valid. As a result, we arrive at
\begin{equation}
\left(\alpha_\mu\partial_\mu +\kappa P\right)N^\pm(x,x')=0 .
 \label{64}
\end{equation}
The vacuum expectation of chronological pairing of operators
(propagator) is defined by the equation
\[
\langle T\Psi_{M}(x)\overline{\Psi}_{ N}(y)\rangle_0=N^c_{
MN}(x-y)
\]
\vspace{-6mm}
\begin{equation} \label{65}
\end{equation}
\vspace{-6mm}
\[
=\theta\left(x_0 -y_0\right)N^+_{MN}(x-y)+\theta\left(y_0
-x_0\right)N^-_{MN}(x-y) ,
\]
where $\theta(x)$ is the theta-function. One obtains from Eq.(65):
\begin{equation}
\langle T\Psi_{M}(x)\overline{\Psi}_{N}(y)\rangle_0 =\left(
\frac{\alpha_\mu\partial_\mu- \kappa \overline{P}
}{\kappa}\right)^2_{MN}D_c (x-y),
 \label{66}
\end{equation}
where the function $D_c (x-y)$ is given by
\begin{equation}
D_c (x-y)=\theta\left(x_0 -y_0\right)D_+(x-y)+\theta\left(y_0
-x_0\right)D_-(x-y) . \label{67}
\end{equation}
Expressions (56), (57), (66) can be used for calculating different
electrodynamics processes involving polarized massless particles.
Taking into consideration the equation \cite{Ahieser}
$\partial_\mu^2D_c (x)=i\delta(x)$, we find
\begin{equation}
\left(\alpha_\mu\partial_\mu+ \kappa P\right)\langle
T\Psi(x)\cdot\overline{\Psi}(y)\rangle_0 =
\frac{i}{\kappa^2}\left(\alpha_\mu\partial_\mu- \kappa
\overline{P}\right)\delta (x-y).
 \label{68}
\end{equation}
It should be noted that propagator (66) includes the contribution
of the spin-0 state.

\section{Conclusion}

We have considered the generalized Maxwell equations that describe
massless fields including spin-0 state (the ghost). The reason for
this is possible cosmological applications or further
investigations of quantum field theory with the indefinite metric.
The ghost state gives the negative contribution to the
Hamiltonian, which requires to introduce an indefinite metric
\cite{Kruglov1}. Quantization of fields and propagator obtained
allow us to make some necessary calculations in the quantum theory
considered.

One can arrive at the classical Maxwell equations by imposing the
condition $\psi_0 (x)=0$ to eliminate the ghost state. As a
result, the gauge invariance will be recovered, longitudinal
states eliminated, and one has only two spin states with helicity
$\pm1$.

One can speculate that such fields with multispin 0 and 1 exist.
Then, we can consider the coupling of these massless fields to the
electromagnetic fields of the form
\begin{equation}
{\cal L}_{int}=g\psi_0 F_{\mu\nu}^2+\lambda\psi_{[\mu\nu]}F_{\mu\nu},
\label{69}
\end{equation}
where $F_{\mu\nu}$ is the strength of the electromagnetic fields. The first term in
the effective Lagrangian (69) is the axion-like interaction of spin-0 state with
the electromagnetic fields \cite{Sikivie}. The interaction (69) may result in the effect of
vacuum berefringence and dichroism \cite{Raffelt}. Now such effects are of the experimental
interest \cite{Zavattini}. Different schemes, leading to vacuum berefringence and dichroism
are discussed in \cite{Ahlers}, \cite{Kruglov3}.

\end{document}